# CATS: Empowering the next generation of rocket scientists through educational flight computers

**Jonas Binz** [1], **Nemanja Stojoski** [1], **Luca Jost** [1]

[1] *Control and Telemetry Systems GmbH, Zug, Switzerland, info@catsystems.io*

**ABSTRACT**

This paper presents an in-depth analysis of the Vega flight computer, and its corresponding ground station developed by CATS, a company producing open-source flight computers and tracking systems tailored for student-made rockets. These flight computers, designed to support rockets reaching altitudes of up to 30 km and possibly higher, play a crucial role in advancing educational rocketry and facilitating hands-on learning experiences in aerospace engineering. As the official sponsor of the European Rocketry Challenge (EuRoC), these flight computers have become integral to the competition, providing reliable and sophisticated telemetry and control capabilities that enhance both safety and performance. The paper delves into the technical specifications and educational impact of these systems, highlighting their contribution to the broader European rocketry programmes. Through comprehensive field data and case studies from the recent European Rocketry Challenge, this study underscores the potential of open-source flight computers to inspire the next generation of aerospace professionals.

## 1. INTRODUCTION

Flight computers are critical components in modern rocketry, serving as the brain of the rocket by managing various control and telemetry functions. These systems are responsible for monitoring the rocket's trajectory, triggering the recovery mechanisms, and ensure the collection and transmission of vital flight data back to ground stations. They may also manage propulsion and stabilization of the rocket. The sophistication of flight computers can significantly contribute to the success and safety of a rocket launch, making them indispensable in both professional and educational rocketry endeavours.

In Europe, student rocketry has seen substantial growth and development over the past decade, driven by a surge in interest in STEM (Science, Technology, Engineering, and Mathematics) education and the increasing accessibility of advanced aerospace technologies[1], [2], [3], [4]. This field has fostered numerous university-led rocketry teams and initiatives, aimed at providing practical, hands-on experience in rocket design, construction, and flight.

A notable platform for these student-led projects is the European Rocketry Challenge, an annual competition that brings together university teams from across Europe to showcase their rocketry skills [5]. The competition emphasizes not only the importance of theoretical knowledge but also highlights the need for practical expertise in areas such as propulsion, aerodynamics, and flight control systems. Participants are required to design, build, and launch rockets capable of reaching altitudes of either 3 km or 10 km depending on their category, presenting a formidable challenge that mirrors real-world aerospace projects.

Central to the success of these student rockets are advanced flight computers, such as those developed by Control and Telemetry Systems GmbH [6]. As the official flight computer system of EuRoC, Control and Telemetry Systems GmbH provides open-source flight computers that are both sophisticated and accessible, enabling student teams to leverage cutting-edge technology in their projects. These flight computers offer a robust platform for monitoring the rocket's flight path, collecting telemetry data, triggering the recovery mechanisms, and ensuring overall mission success.

The integration of open-source flight computers into student rocketry projects not only enhances the technical capabilities of these rockets but also democratizes access to high-quality aerospace technology. This democratization is crucial for fostering innovation and providing students with the tools they need to excel in aerospace engineering. By collaborating with academic institutions and supporting initiatives like EuRoC, Control and Telemetry Systems GmbH is playing a pivotal role in shaping the future of European student rocketry, inspiring a new generation of aerospace professionals.

## 2. THE VEGA FLIGHT COMPUTER

The CATS Vega flight computer, along with its associated ground station and configurator software, forms a comprehensive system designed for use in student-made rockets. This section details the hardware components, setup procedures, and operational principles used to implement and utilize the CATS Vega system for rocketry projects.

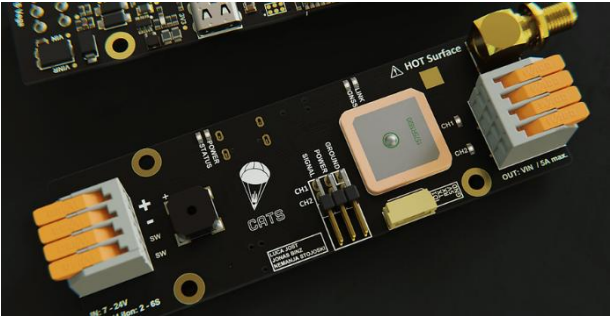

*Figure 1: Vega Flight Computer*

### 2.1. Hardware

The CATS Vega flight computer is a compact, lightweight unit with dimensions of 100 x 33 x 21 mm and a weight of 25 grams. It operates within an input voltage range of 7-24V and consumes 100mA of power. The device includes two pyro channels, two servo channels, one general-purpose I/O port, and an additional UART interface. It is built around an STM32F4 microcontroller, with 16MB of flash memory. Sensor integration includes an LSM6DSO32 IMU and an MS5607 barometer. The telemetry system operates in the ISM 2.4GHz band, with power up to 1W, providing a tested range of over 10km at 100mW.

Key hardware features include a manual switch port, battery port, buzzer for readiness beeps, status LEDs, USB connector, test button, servo connector, telemetry LEDs, low-level I/O and UART connector, and pyro channel connectors. An antenna connector is also included for telemetry data transmission.

### 2.2. Software Overview

The Vega open-source embedded software consists of many different components (Figure 2). An IMU and Barometer are read out with 100 Hz, feeding the state estimation which implements a simple 1 DoF Kalman filter. The output of the Kalman filter as well as the sensor data is fed into a Finite State Machine (FSM) shown in Figure 3. Whenever a transition in the finite state machine is taken, an event is fired. On this event, the user can tie any action, including timers, GPIO outputs, pyro outputs and servo channels, triggering any type of connected hardware. All this information is recorded on the onboard flash chip through the recorder. The same information is also sent with 10 Hz to the telemetry chip which sends the data using the LoRa protocol [7].

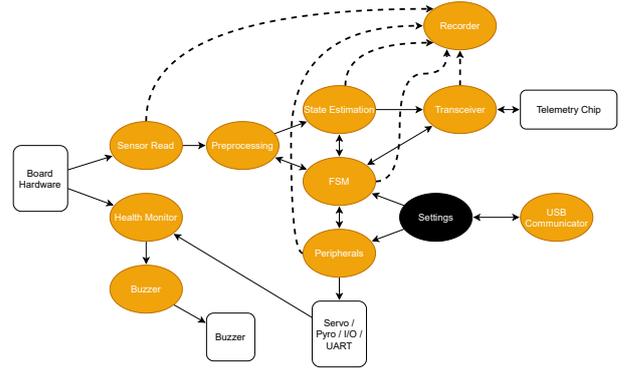

*Figure 2: Software Overview*

### 2.3. Kalman Filter

Here we quickly derive the Kalman filter [8] used in the Vega flight computer. We assume the state and the noise to be

$$x(t) = \begin{pmatrix} h(t) \\ v(t) \\ a_0(t) \end{pmatrix} \qquad v(t) = \begin{pmatrix} v_1(t) \\ v_2(t) \end{pmatrix} \quad (1)$$

where $h$ is the height above ground level, v is the vertical velocity and $a_0$ is the estimated offset of the measured vertical acceleration. $v_1$ is the noise applied on the vertical acceleration and $v_2$ is the noise variable used to estimate the linear acceleration offset. Additionally, we need to define the input to the system, which, in this derivation is $u(t)$ being the linear acceleration measured in z direction.

$$\dot{x}(t) = \begin{pmatrix} v(t) \\ a(t) \\ \dot{a}_0(t) \end{pmatrix} = Ax(t) + Bu(t) + Gv(t) \quad (2)$$

$$= \begin{pmatrix} 0 & 1 & 0 \\ 0 & 0 & 1 \\ 0 & 0 & 0 \end{pmatrix} \begin{pmatrix} h(t) \\ v(t) \\ a_0(t) \end{pmatrix} + \begin{pmatrix} 0 \\ 1 \\ 0 \end{pmatrix} u(t) + \begin{pmatrix} 0 & 0 \\ 1 & 0 \\ 0 & 1 \end{pmatrix} v$$

this is discretized using a first order approximation

$$A_d = e^{AT_s} \quad B_d = \int_0^{T_s} e^{A_d t} \, \delta t \, B \quad (3)$$

$$G_d = \int_0^{T_s} e^{A_d t} \, \delta t \, G \quad (4)$$

giving the system

$$x(k+1) = A_d x(k) + B_d u(k) + G_d v(k) \quad (5)$$

The process measurement noise matrix becomes

$$Q(k) = \begin{pmatrix} Q_{acc}(k) & 0 \\ 0 & Q_{acc_0}(k) \end{pmatrix} \quad (6)$$

The measurement step assumes, that the height is already calculated from the barometric pressure. For this we use the standard barometric pressure formula for a linear temperature gradient.

$$z(k) = h_{meas}(k) + w(k)$$

$$= \left(\frac{p(k)^{\frac{1}{5.275}}}{p_0} - 1\right)\frac{T_0 + 273.15}{L} - h_0 + w(k) \quad (7)$$

with $L = -0.0065$, $T_0 = 15\,C$, $p_0 = 101250\,Pa$, $h_0 = calibrated\ altitude\ above\ sea\ level$ and $\omega$ being the measurement noise. The measurement function becomes, very easily

$$z(k) = H\,x(k) = (1 \quad 0 \quad 0) \begin{pmatrix} h(k) \\ v(k) \\ a_0(k) \end{pmatrix} \quad (8)$$

The measurement noise matrix becomes then a scalar $R(k) = R_{height}$. This can then be gain-scheduled according to the finite state machine to reduce pressure spikes during the thrusting phase, making the algorithm more robust against transonic waves.

### 2.4. Finite State Machine

The FSM defines the state of the flight computer. In the *Calibration* phase, no action can be taken. The flight computer transitions into the *Ready* phase when no IMU action is detected for 10 seconds. In this state, the flight computer is ready to detect liftoff. Liftoff is detected if an acceleration bigger than the user defined value is detected. This allows the user to fine-tune the liftoff detection based on the motor used. In the *Thrusting* phase, the Kalman filter trusts the IMU the most as pressure readings are not very reliable during this phase (particularly in the transonic regime). *Coasting* is detected when the acceleration in z direction falls below 0, indicating that the motor stopped thrusting. At this point, the Kalman filter slowly increases the weight of the barometric pressure reading depending on the velocity, trusting the barometric pressure more with decreasing velocity. Apogee is detected when the velocity of the rocket, estimated by the Kalman filter falls below 0 m/s for a defined time. This is usually the event connected to the drogue parachute deployment mechanism. In the next phase, the Kalman filter ignores IMU readings as the dangling of the rocket on the parachute makes those reading unusable for z-direction measurements. The transition to *Main* is achieved when the estimated height above ground level is below the user defined threshold. This transition is typically connected with triggering the main parachute. Lastly, *Touchdown* is detected when no IMU action is detected for a certain amount of time.

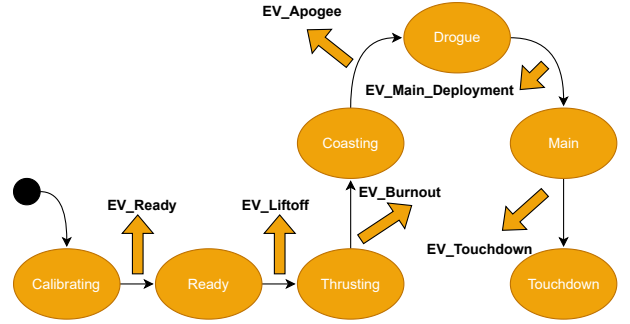

*Figure 3: Finite State Machine of the Vega*

### 2.5. Flight Preparation

Flight preparation involves mounting the CATS Vega flight computer within the rocket and ensuring all connections are secure. The battery, switch, and actuators are connected according to the hardware specifications. The flight computer is configured according to the specifications of the student rocket, including triggering of pyro channels, servo motors or solenoid valves according to the flight phases depicted in sec. 2.4. The flight computer can be mounted in any direction inside the rocket and needs a RF-transparent structure for good GNSS reception and telemetry.

### 2.6. Telemetry

The data is then transmitted from the Vega flight computer to the corresponding ground station. The system is based on 2.4 GHz LoRa and utilizes Frequency-Hopping Spread Spectrum (FHSS) for transmission [9]. key benefit of FHSS is that it makes transmissions much more resistant to interference and more difficult to intercept. Additionally, it allows more devices on the same frequency band with little or no impact on the link quality.

The hopping pattern is defined with a link phrase. This link phrase is hashed using a CRC32 algorithm, and the resulting hash value is then used as the seed for a pseudo-random number generator. The generator is run 20 times, defining the hopping pattern. With this method, a link phrase always generates the same hopping pattern. Both the transmitter and receiver require the same link phrase for the transmission to work.

The receiver waits on the first frequency until a sync packet is received. The sync packet contains the link CRC (Cyclic Redundancy Check) to identify the transmission source. If the remote CRC matches the local CRC, the receiver hops to the subsequent frequency, waiting for data. Each data package contains a checksum to validate the contents. The time between packages is measured and used to jump to the next frequency when no package was received in the estimated time frame. A total of 30 hops can be performed without receiving a package before the synchronization is lost. On connection loss, the receiver returns to the first frequency.

### 2.7. Flight Execution

During flight, the CATS Vega flight computer operates based on a finite state machine with configurable actions. These actions include detecting liftoff through acceleration thresholds, deploying parachutes at specific altitudes, and triggering timers for various flight events. The telemetry system transmits real-time flight data to the ground station, allowing for live monitoring and post-flight analysis.

### 2.8. Post-Flight Analysis

After the flight, data from the CATS Vega flight computer is downloaded and analysed using the configurator software. The data includes sensor readings, event logs, and telemetry information, which can be visualized to assess the flight performance and diagnose any issues. Software updates can also be applied to improve system functionality and address any bugs or issues identified during the flight.

## 3. GROUND STATION

The ground station, a crucial counterpart to the CATS Vega flight computer. It is designed to receive data from the Vega board and send commands to it. This setup enables real-time tracking of the rocket's position and velocity, along with monitoring its health and other essential parameters. The ground station operates around an ESP32-S2 microcontroller and features a transflective display for optimal readability even under bright sunlight. It includes an onboard flash memory capable of storing up to 1MB of data, which suffices for tracking over an hour of flight data.

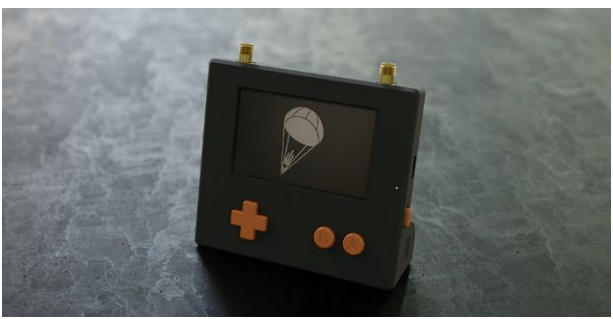

*Figure 4: Ground station*

### 3.1. Hardware Specifications

The ground station incorporates an ESP32-S2 microcontroller, 4MB of flash memory, and is powered by a Li-Ion 18650 battery. It consumes 60mA of power and can be charged at a current of 500mA via USB. The display used is an LS027B7DH01 screen, and it includes two SX1280 radios tested for a range of up to 10km at 100mW. Additionally, it is equipped with a GNSS module ATGM336H-5N.

### 3.2. Operation and Interface

The ground station's interface is navigated using a joystick and two buttons (A and B). The live data screen displays all the received data from the Vega flight computer, including rocket altitude, vertical velocity, GNSS coordinates, battery voltage, pyro continuity, and errors. Information about the telemetry link, such as package age, signal-to-noise ratio (SNR), link quality (LQ), and received signal strength indication (RSSI), is also displayed.

### 3.3. Recovery

The recovery window helps in locating the rocket post-flight using the last known GNSS coordinates. The ground station's onboard sensor suite calculates the distance and direction to the rocket. For accurate direction tracking, the device's compass must be calibrated outdoors near the launch site, free from large metallic objects.

### 3.4. Charging and Power Management

The ground station is powered by a Li-Ion 18650 battery, providing more than 8 hours of operation when fully charged. The battery can be charged through the USB port, taking up to 6 hours to fully charge at a current of 500mA. An LED indicator next to the USB port lights up during charging and turns off once the battery is fully charged. The battery is replaceable by removing the cover on the back of the ground station.

### 3.5. Data Logging and Software Updates

The ground station can be connected to a computer, recognized as a mass storage device, allowing users to transfer recorded logs, stored in .csv format, to their preferred location. For software updates, the device can enter DFU mode by selecting the bootloader setting in the settings panel or manually by using hardware buttons after opening the casing. The firmware files, with a .UF2 extension, can be dragged and dropped into the SAOLA1RBOOT mass storage device that appears on the computer.

### 3.6. Telemetry Modes

The ground station supports two telemetry modes: single and dual. In single mode, it tracks one Vega computer using both antennas to receive more data. In dual mode, it can track two Vega computers simultaneously, useful for scenarios where a section of the rocket separates and needs to be tracked in addition to the main body. Using a combination of directional and omnidirectional antennas in diversity mode is recommended for optimal performance.

### 3.7. Menu Navigation

The ground station's menus include live data, recovery, testing, flight information, sensors, and settings. The sensors tab displays raw data from the onboard IMU, magnetometer, and GNSS module, and allows magnetometer calibration. The settings menu provides options for time zone configuration, logging preferences, software version display, bootloader initiation for updates, telemetry mode selection, and link phrase setup for connecting to Vega computers.

This comprehensive setup ensures robust tracking, data logging, and command functionalities for the CATS Vega flight computer, facilitating detailed monitoring and recovery of student-made rockets during high-altitude missions.

### 4.1 EuRoC Example Flight

Through the logging capabilities of the Vega flight computer, a multitude of sensing data can be acquired. Here, we show an example flight of the rocket from RED (Rocket Experiment Division) from Instituto Superior Técnico, Portugal. The interesting aspect of this rocket flight is that the team flew an airbrake system, controlled to reach an altitude of 3000 m. The max altitude reached was 3014 m according to the Vega flight computer. For the recovery, the rocket used a dual stage recovery system where both parachutes are fired using a pyro charge.

Looking at the altitude measurements from the barometer, one can see that the ascent was nominal with some bumps in the barometric readings after burnout. This shows the effect of the airbrakes onto the rocket. In amateur rocketry this can become a big problem for student researched and developed (SRAD) flight computers. Detecting apogee is usually done by checking the velocity of the rocket: if it is negative, apogee is reached. However, with those kind of readings,

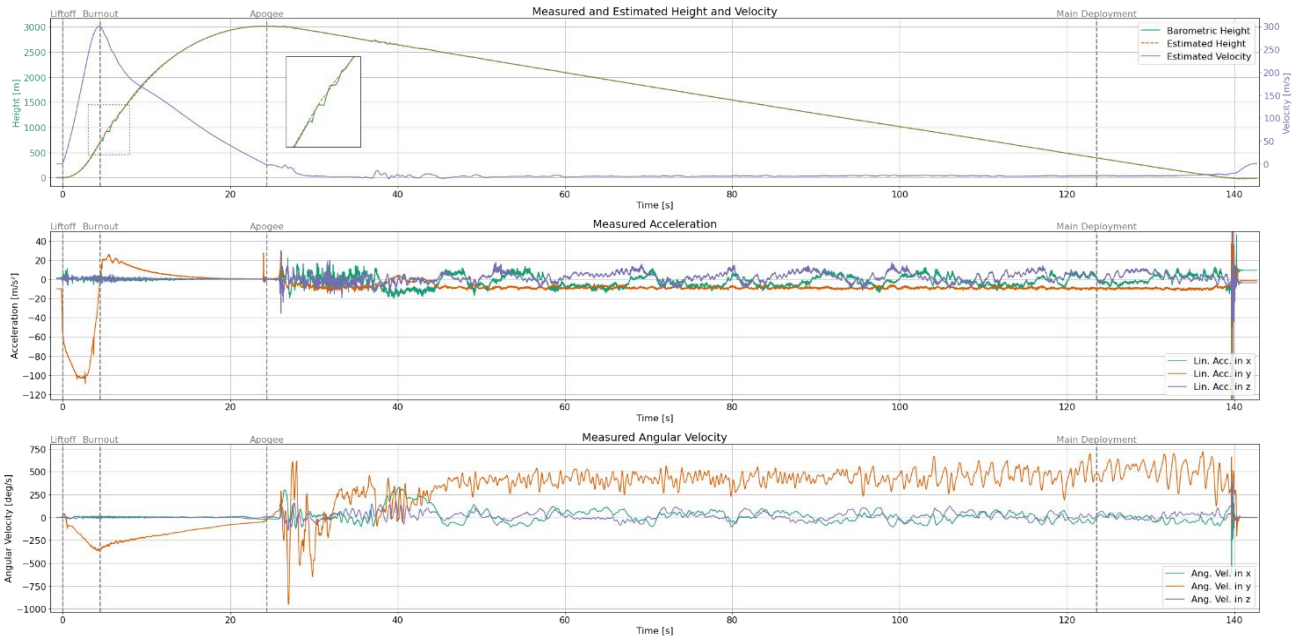

*Figure 5: Flight data from the Vega flight computer from team RED during EuRoC 2023*

### 4. EUROC 2023

During EuRoC 2023 a total of 18 flight computers were flown on 13 different rockets. All 18 flight computers detected the flight phases of the rocket flight, properly triggering the drogue as well as the main parachute. The data of all those flights can be found under [10].
Of those 13 rockets, 12 aimed for an altitude of 3 km while one rocket aimed for 9 km. The average telemetry reception, across all rockets, was 65.2 % (σ = 8.5) of all sent packets. This corresponds to an average reception rate of 6.5 packets per second.

premature apogee might be detected, depending on the velocity computation. However, by fusing barometric pressure and IMU readings using a filtering approach, this risk can be mitigated as shown in the estimated velocity. The maximum velocity, estimated by the pressure and IMU measurements, shows a peak around 300 m/s. At apogee, the flight computer fires the pyro charges, the recovery mechanism of the rocket for the drogue parachute. The chute is ejected, and the rocket begins its descent with an estimated downward speed of 30 m/s. At 400 m above ground level the flight computer fires the second charges of pyros. However, due to a recovery mechanism malfunction, the parachute is not ejected. This can also be seen in the velocity readings, there is no decrease.

Another interesting measure is the linear acceleration in the z direction. During the thrusting phase a peak acceleration of 100 m/s$^2$ is measured. After burnout, the effect of the airbrakes can also be seen on the linear acceleration readings. In particular, it may be interesting to combine those readings with the airbrake extension to compute the effective drag which the airbrakes exert on the rocket. From the IMU readings it also becomes clear that the secondary flight computer fired the main chute. At 24 seconds after liftoff, 0.35 seconds before the Vega flight computer fires the apogee event, a bump in the linear acceleration is seen. This is most probably from the pyro charge being ignited. The parachute starts to slow down the rocket 1.9 seconds later, at 25.9 seconds after liftoff. This can also be seen in the IMU readings.

The rocket was rather stable during ascent, as can be seen in the gyroscope readings. A certain roll during ascent is expected and necessary for a stable trajectory. After apogee, the rocket however still had a lot of roll, spinning on its own axis while it was descending on the parachute.

## 5. CONCLUSION

The development and deployment of the CATS Vega flight computer system marks a significant milestone in the advancement of educational rocketry. Through its integration into student-led projects and competitions such as the European Rocketry Challenge (EuRoC), the Vega system has proven to be an invaluable tool improving the safety and tracking capabilities of those rockets. By offering a reliable and robust platform for telemetry, data collection, and mission control, the system not only enhances the technical sophistication of student rockets but also plays a crucial role in ensuring their success during critical phases of flight.

Moreover, the open-source nature of the Vega system embodies the democratization of access to advanced aerospace technology. This approach empowers students and educators by providing them with high-quality tools that were once the exclusive domain of well-funded research institutions and commercial entities. As a result, the Vega system not only contributes to the growth of individual projects but also supports the broader objective of nurturing the next generation of aerospace professionals. It serves as both an educational resource and a catalyst for innovation, inspiring students to push the boundaries of their knowledge and capabilities.